\documentclass[12pt]{article}
\usepackage{epsfig}
\usepackage{cite}
\usepackage{mcite}
\usepackage{url}

\usepackage{array,tabularx,epsfig,mathrsfs,graphicx,rotating}
\usepackage{ifthen}
\usepackage{fancybox}
\usepackage{amsfonts}

\newcommand{\beq}{\begin{equation}}
\newcommand{\eeq}{\end{equation}}

\chardef\til=126

\newcommand{\gev}{{\,\mathrm{GeV}}}

\begin{document}

\clearpage
\pagestyle{empty}
\setcounter{footnote}{0}\setcounter{page}{0}%
\thispagestyle{empty}\pagestyle{plain}\pagenumbering{arabic}%

\hfill  ANL-HEP-PR-10-2 
 
\hfill January 10,  2010

\hfill May 20, 2010 (updated)

\vspace{2.0cm}

\begin{center}

{\Large\bf
Searches for TeV-scale  particles at the LHC using jet shapes 
\\[-1cm] }

\vspace{2.5cm}

{\large S.V.~Chekanov and J.~Proudfoot 

\vspace{0.5cm}
\itemsep=-1mm
\normalsize
\small
HEP Division, Argonne National Laboratory,
9700 S.Cass Avenue, \\ 
Argonne, IL 60439
USA
}

\normalsize
\vspace{1.0cm}


\vspace{0.5cm}
\begin{abstract}
New particles at the TeV scale can decay hadronically
with strongly collimated jets, thus the standard reconstruction methods
based on invariant-masses of well-separated jets can fail.  We discuss how to identify 
such particles in $pp$ collisions at the LHC
using simple jet shapes which help to reduce the contribution 
of QCD-induced events. We focus on 
a rather generic example $X\to t\bar{t}\to hadrons$,
with $X$ being a heavy particle, 
but the approach is well suited for reconstruction of other decay channels
characterized by a cascade decay of known states.  

\end{abstract}

\end{center}

\newpage
\setcounter{page}{1}


\section{Introduction}

Proton-proton collisions at the LHC will allow physicists to search for
new TeV-mass particles that can corroborate or disprove current theories such as the
Standard Model (SM).
One promising path to discoveries at the LHC
is through model-independent searches
in which events can be
classified in exclusive classes according to the number of identified high-$p_T$ objects,
such as jets and leptons.  
Such event classes can be analyzed by reconstructing event rates,
invariant masses and other event characteristics.
However, heavy particles with masses close to the TeV scale 
can decay into more than two states which undergo   
a significant Lorentz boost.
This leads to partial or complete overlap of the decay products 
which cannot be reconstructed as separate objects. In the case of jets, this
closes the opportunity
of bump hunting in invariant-mass
spectra using individual jets since the event signatures will be indistinguishable from 
that of the standard QCD-induced events.
 
One approach to tackle this problem
is to use jet shapes (see 
recent 
publications \cite{Agashe:2006hk,Lillie:2007yh,Butterworth:2007ke,Almeida:2008tp,Almeida:2008yp,
Kaplan:2008ie,Brooijmans:2008,Butterworth:2009qa,Ellis:2009su,ATL-PHYS-PUB-2009-081,CMS-PAS-JME-09-001}).  
One hopes that its distinct 
characteristics can be useful for reduction of the overwhelming rate
of conventional QCD jets, thus opening the path to a direct observation of new  states.  

In this paper we will discuss how to reconstruct a rather generic decay $X\to t\bar{t}$,
(with $X$ being a particle at the TeV scale) using  jet-shape variables. 
The decay channel involving the production of top quarks 
is particularly important since they are expected to be strongly
coupled to the electroweak symmetry breaking sector, thus the channel has a large potential 
for discovery of new states at the LHC. 
Examples of high-mass resonances decaying predominantly to $t\bar{t}$  
have recently been discussed 
in \cite{Matsumoto:2006ws,Agashe:2006hk,Fitzpatrick:2007qr,Lillie:2007yh}. 
Additional channels include technicolor models, 
strong electroweak symmetry breaking models, topcolor, SM Higgs, MSSM Higgs etc.
(see \cite{Brooijmans:2008se} for a review).

Recently, a detailed discussion of how to reject QCD background using monojets
(jets that fully contain decay products of top quarks) has been discussed within
the model with the lightest Kaluza-Klein (KK) excitation of the gluon $g^{(1)}\to t\bar{t}$ in 
Randall-Sundrum scenario \cite{Agashe:2006hk,Lillie:2007yh,Thaler:2008ju}. It was pointed out 
that jet substructure can  be useful
to increase the signal-to-background ratio
since more conventional approaches will fail
due to the large overlap of decay products.
For example,  $b$-tagging may be  inefficient 
because of a large boost of $b$-quarks from top decays
and, therefore,
a significant collimation of the decay products from the $b$ decays. 
It was concluded that extraction of the KK signal requires a background
rejection of about a factor 10 \cite{Lillie:2007yh}.

In this paper we will extend the studies presented in Ref.~\cite{Lillie:2007yh}, 
focusing on jet shapes 
for rejection of QCD background in the hadronic decay $X\to t\bar{t}$.
We will show that using only jet-shape variables, 
one can easily reduce 
the contribution
to the jet-jet invariant masses from the standard QCD by more than one hundred,
with only a factor three reduction for the signal events.  
Thus we  confirm a rather good perspective on searches for heavy-mass
particles in  the $t\bar{t}$ decay channel,
or any other similar decay channel with a cascade decay of known states. 
In addition, we extend the reconstruction aspect of the previous 
paper \cite{Lillie:2007yh} in several areas:
(a) We will consider the anti-$k_T$ algorithm \cite{Cacciari:2008gp} which is 
used  for the default jet reconstruction at the ATLAS experiment;  
(b) we will study the jet-shape variables 
and calculate correlations between them;   
(c) using the jet shapes for background rejection,  we 
will estimate a minimal cross section necessary for $6\sigma$ observation of a 
new TeV-scale particle 
using the dijet invariant-mass
signatures and $200$ pb$^{-1}$ of $pp$ collisions at $10$ TeV.
We argue that the approach allows to obtain a similar   
discovery-level cross section that was previously anticipated for 10 fb$^{-1}$ of 
$pp$ collision
at $14$ TeV \cite{Beneke:2000hk}.

\section{Jet shapes for searches of TeV-scale particles} 
\label{sec1}

The idea of reconstruction of top quarks using monojets,
i.e. collecting all 
energy deposits  in a large cone around the top-quark direction,
has previously  been discussed  (see, for example, \cite{Beneke:2000hk}).
A new approach to this problem is to use jet shapes 
to reject QCD background as recently discussed in several publications
(see, for example, the most recent 
papers \cite{Butterworth:2007ke,Almeida:2008yp,Almeida:2008tp,Kaplan:2008ie,Brooijmans:2008,Ellis:2009su, Butterworth:2009qa,ATL-PHYS-PUB-2009-081, CMS-PAS-JME-09-001}).

\begin{figure}[htp]
\begin{center}
\mbox{\epsfig{file=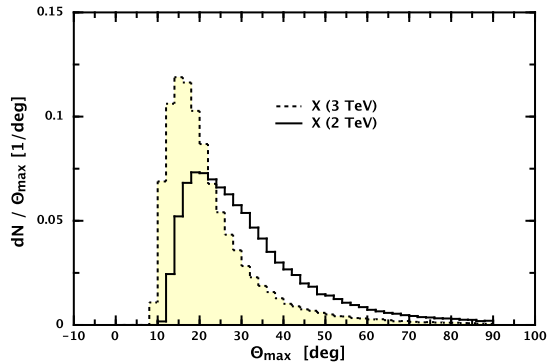,width=8cm}}
\caption
{
A Monte Carlo simulation of the probability
distribution for the maximum angle between quarks ($q,b$) coming 
from the parent top quark in the decay $X\to t\bar{t}$, with
$t\to W b \to q \bar{q} b$.
The simulation consists of a calculation of the decay angles
in the center-of-mass frame of the daughter particles and
subsequent Lorentz boost to the laboratory  frame.
The original momentum of the particle $X$ was set to 10 GeV,
but the distributions shown depend only on its mass,
which was set to 3 TeV (dashed line) or 2 TeV (continues line).
}
\label{zprime}
\end{center}
\end{figure}

Before going into a detailed description of the jet-shape approach, we recall 
a few basic kinematic features when dealing with decays of heavy particles. If a TeV-mass particle $X$
decays into several known particles without subsequent cascade decay, 
then their decay products are spatially separated.
For example, a heavy graviton decaying into $\gamma\gamma$  should not cause
reconstruction problems since the photons are produced back-to-back in the laboratory frame.
The situation is dramatically different for a cascade-type of decays; due to a significant Lorentz
boost, the decay products can easily overlap. This is illustrated in Fig.~\ref{zprime}
which shows the largest angular separation between three quarks originating from the same
parent top quark
in the reaction  $X\to t\bar{t} \to W^+b_1 W^-b_2$, with subsequent
decays of the $W$ bosons into two quarks, $W\to q_1 q_2$.
The figure was obtained using a simple Monte Carlo simulation which consists of
two steps: 1) a calculation of decays
in  the center-of-mass frame of daughter particles, assuming flat 
distributions of the azimuthal and polar angles in this frame; 
2) performing a  Lorentz boost of daughter particles
to the laboratory  frame.
The original momentum of the $X$ particle does not affect the
angles between $q_i$ quarks and the top quark, while the mass of the state $X$
has a significant impact on the angular distribution.   
A similar distribution can be obtained for simpler decays, 
such as $X\to W^+ W^- \to 4\>q$.

The above observation has a direct implication for experimental searches of heavy particles.
The $k_T$ jet algorithm \cite{Catani:1993hr,*Ellis:1993tq} with the size
$R=0.4$ collects 
energy deposits into a single jet if angular separation between
them is not larger than $21^o$;
thus two jets cannot be resolved as two separate objects if 
jet centers are separated by an angle smaller than this angle.
In the case of the size $R=0.6$, 
the minimal angular separation is $31^o$.
A similar conclusion applies to cone algorithms\footnote{This is not an exact statement, 
since clustering for the $k_T$ and for the conventional cone algorithm are done differently 
assuming the same input jet-size parameter.}.
Figure~\ref{zprime} shows that a significant fraction of events contain
decay products
separated by angles less than $21^o$ ($31^o$). In the case of $3$ TeV particles,
the largest fraction of jets will overlap and will not be reconstructed as separate objects.
A similar conclusion was obtained in \cite{Lillie:2007yh}.

As mentioned before, we will use the anti-$k_T$ 
version  \cite{Cacciari:2008gp} of the popular $k_T$ algorithm \cite{Catani:1993hr,*Ellis:1993tq}.
Our strategy for identification of jets  from 
top quarks produced by  TeV-scale particles
will be based on three jet-shape variables.
In the case of fully overlapping jets, one should expect two energetic jets, each of which
is due to the hadronic  decay of a top quark.
Thus we will be interested in the shapes of two jets 
leading in transverse momentum ($p_T$).
The following jet-shape variables will be considered:
the jet mass, the jet width and the eccentricity parameter.
The jet mass calculated as the mass of all  constituents inside a jet,
and the jet width and jet eccentricity are considered below:

\subsection{Jet width}
 
The jet width is  defined as the $p_T$ weighted
sum of jet constituents falling into the ring $dR$ in pseudorapidity
$\eta$ and azimuthal angle $\phi$ centered with respect to the jet axis.  Mathematically,  
it can be written as :
\begin{equation}   
\frac{\sum_{i=1}^{N} dR^{i} p_{T}^{i}} {\sum_{i}^{N} p_{T}^{i}}, 
\label{width}
\end{equation}
where $dR^i=\sqrt{d\phi^2_i + d\eta^2_i}$ defines a ring positioned at the jet center,
with  $d\phi_i$ and $d\eta_i$ being distances in the azimuthal angle and pseudorapidity
of a jet constituent $i$ with a transverse momentum $p_{T}^{i}$ from the jet center.
The sum runs over all particles $N$ inside the jet.

\subsection{Jet eccentricity}
The jet eccentricity, ECC, is defined as
$1-v_{max} / v_{min}$, where $v_{max}$ ($v_{min}$) is 
the maximal (minimal) values of variances of jet constituents 
along the principle (minor) axis.
The calculation consists of a few steps: First,
for each jet constituent $i$ 
with the energy $e_i$, 
the energy-weighted  centers in $\eta$ and $\phi$ are calculated as:

\begin{equation}   
\bar{\phi}= \sum_{i=1}^N  d\phi_i\>  e_i / \sum_{i=1}^N  e_i, \qquad 
\bar{\eta}= \sum_{i=1}^N  d\eta_i\>  e_i / \sum_{i=1}^N  e_i, 
\label{eec}
\end{equation}
where $d\phi_i$ and $d\eta_i$ are defined as for the jet width in
Eq.~(\ref{width}).
The differences from the  energy-weighted  centers are given by 
$\Delta \eta_i = \eta_{i} - \bar{\eta}$  and $\Delta \phi_i = \phi_{i} - \bar{\phi}$.
Then, the standard principle-component analysis \cite{PCA} was performed  
to determine the axis along which 
the energy-weighted                    
variance of jet constituents is maximized (principle axis), 
while the orthogonal axis was defined as the minor axis.
The rotation angle $\theta$ of the ellipse populated by hadrons 
inside a jet is given by:
\begin{equation} 
\tan(2 \theta)   =  
\frac{2\sum_{i=1}^N   e_i \Delta\eta_i \Delta\phi_i}{ \sum_{i=1}^N e_i ( \Delta\phi_i^2 -  \Delta\eta_i^2)},  
\label{eec1}
\end{equation}
while the angle of the orthogonal axis is shifted by $\pi/2$ with respect to $\theta$. 
The energy-weighted variances $v_1$ and $v_2$ for the axes are calculated as:
\begin{equation} 
\begin{array}{lc}
v_1 = N^{-1}\cdot \sum_{i=1}^{N} e_i\> (\cos(\theta) \Delta\eta_i - \sin(\theta) \Delta\phi_i)^2, \\ 
v_2 = N^{-1}\cdot \sum_{i=1}^{N} e_i\> (\sin(\theta)  \Delta\eta_i + \cos(\theta) \Delta\phi_i)^2.  
\label{eec2}
\end{array}
\end{equation}
Finally, the largest value of the variance is assigned to  $v_{max}$,
while the smallest to $v_{min}$. 
The jet eccentricity ranges 
from 0 (for perfectly circular jets) to 1 (for
an infinitely elongated jet shape).
This parameter is similar to 
$\mathrm{det}\> S^{\bot}$  \cite{Thaler:2008ju} discussed within the context   
of possible strategies to identify boosted top quarks. A similar
parameter (the so-called "planar flow" parameter) was also  
discussed in Ref.~\cite{Almeida:2008yp}.

We did not use the jet substructure variables, such as YSplitter values~\cite{Butterworth:2007ke},
which define scales at which a single $k_T$ jet flips into two and three subjets.
For the anti-$k_T$ algorithm, these  
parameters are expected to be sensitive to soft subjets and thus can be strongly 
affected by pileup and jet-reconstruction effects.

\subsection{Studies of correlations}
The analysis of the jet shapes was performed using the PYTHIA Monte Carlo model \cite{Sjostrand:2006za}
included in the RunMC package~\cite{runmc} 
which interfaces FORTRAN Monte Carlo models with ROOT~\cite{root} and other C++ libraries.
Jets and their shapes were reconstructed using the FastJet package~\cite{fastjet}. 

\begin{table}
\begin{center}
\begin{tabular}{|c|c|c|c|}
\hline
      & mass & width & ECC  \\
\hline
mass  & 1    & 0.88 (0.78) & 0.32 (-0.10)  \\
width & 0.88 (0.78)  & 1   & 0.34 (-0.11)  \\
ECC   & 0.32 (-0.10)  & 0.34 (-0.11) & 1   \\
\hline
\end{tabular}
\caption{Pearson's correlation coefficients for the jet-shape variables calculated for
inclusive $pp$ events and events with
$Z' \to t\bar{t} \to 6 q$, where $Z'$ has the mass 2 TeV
(shown in parenthesis).
In both cases the events are generated with the PYTHIA model. The correlation coefficients
are defined as the covariance between $i$th and $j$th jet-shape variable,
divided by the products of the corresponding standard deviations.
In the case of fully correlated variables, the correlation coefficient is one, while
it is zero in the case of uncorrelated variables. The correlation matrix
was obtained for anti-$k_T$ jets with transverse momenta larger than $500\gev$.
}
\label{table1}
\end{center}
\end{table}

One should note that the jet shape variables are not totally independent of each other.
This was checked by running the PYTHIA Monte Carlo model
for fully inclusive $pp$ production at $10$ TeV and reconstructing the jet shapes for
leading jets in $p_T$.
A minimum jet transverse momentum 500 GeV was required.
Table~\ref{table1} shows the Pearson's correlation coefficients  used to estimate
the strength of the linear relationship between the jet-shape variables.
We remind the reader 
that if two observables are strongly correlated, the corresponding correlation coefficient is close
to unity.  Table~\ref{table1} shows that jet widths correlate with masses.
The jet eccentricity  weakly correlates with other variables, which
implies that an additional rejection can be obtained when using this variable.
This conclusion also holds for fully hadronic decays of 
$Z' \to t\bar{t} \to 6 q$ where $Z'$ has the mass 2 TeV, 
although now some variables show a weak anti-correlation.

\section{Searches using jet shapes}

As stated earlier, we are interested in the jet shapes for the decay
$X\to t\bar{t} \to W^+b_1 W^-b_2$, with  $X$ being a particle 
with a mass close to the TeV scale and $W\to q_1 q_2$. 
We will attempt  
to reconstruct the mass of the state $X$ but, more importantly, we will try to   
disentangle the hadronic final-state signatures of this decay from those
of the standard QCD jets using the jet shapes defined in Section~\ref{sec1}.

\begin{figure}[htp]
\begin{center}
\mbox{\epsfig{file=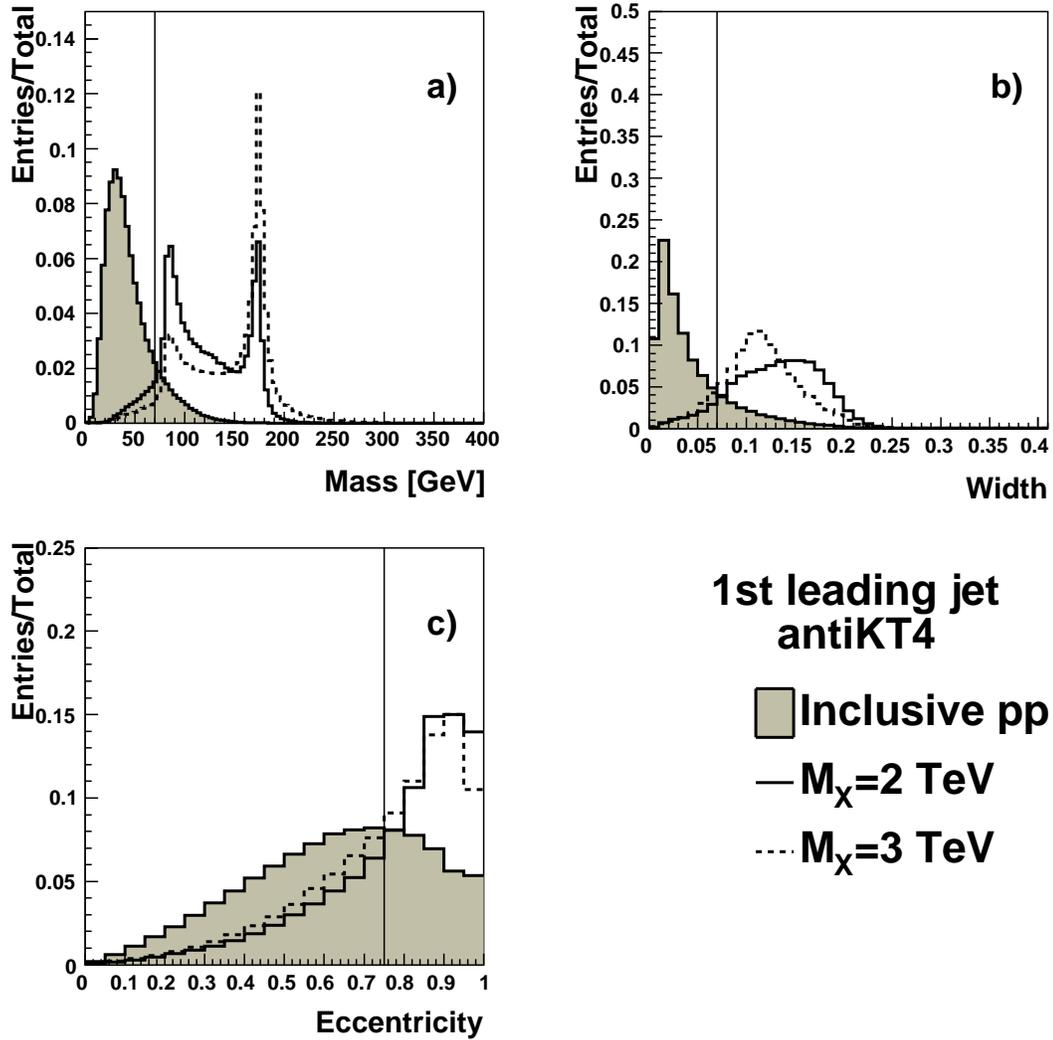,width=15cm}}
\caption
{
Jet shape variables for the leading jet in $p_T$ 
for inclusive
$pp$ collisions (filled histograms) simulated with the PYTHIA model.
Also shown are the shape variables
for $X\to t\bar{t} \to W^+b_1 W^-b_2$, with $W$ bosons hadronically decaying into two jets.
The state $X$ was simulated using a $Z'$ particle
with the mass of $2$ and $3$ TeV (solid and dashed lines, respectively).
Events were selected with at least one jet with $p_T>500$ GeV using the 
anti-$k_T$  jet algorithm.
The vertical lines show the cuts applied to reject inclusive QCD events. 
}
\label{jetkt4a}
\end{center}
\end{figure}

\begin{figure}[htp]
\begin{center}
\mbox{\epsfig{file=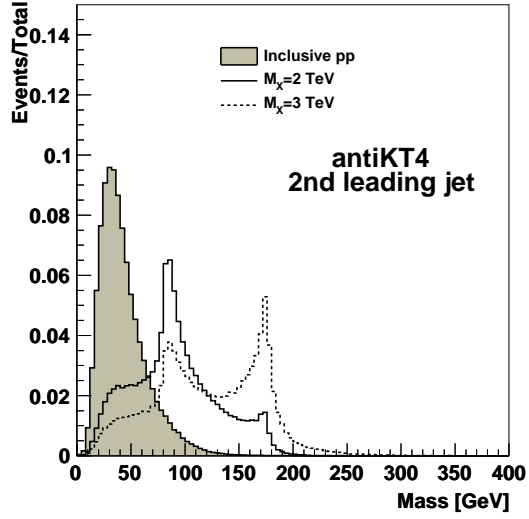,width=8cm}}
\caption
{
The jet masses for the second leading jets in $p_T$  
reconstructed using the anti-$k_T$ jet algorithm with $R=0.4$.
}
\label{jetkt4b}
\end{center}
\end{figure}

\begin{figure}[htp]
\begin{center}
\mbox{\epsfig{file=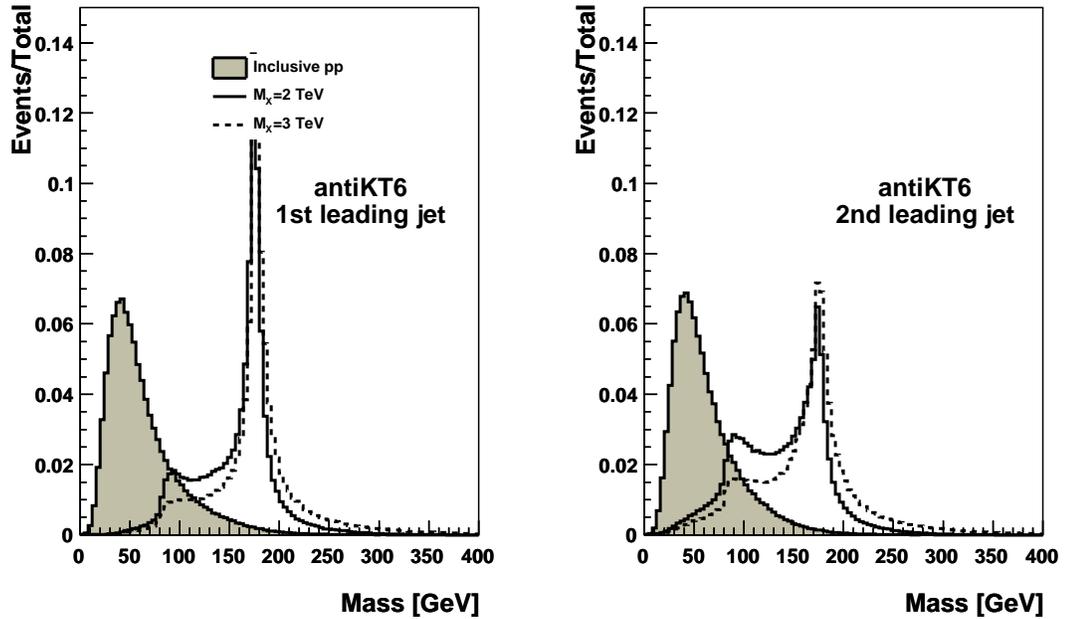,width=15cm}}
\caption
{
The jet masses for the first (left) and second (right) leading jets in $p_T$ 
reconstructed using the anti-$k_T$ jet algorithm with $R=0.6$. 
}
\label{jetkt6}
\end{center}
\end{figure}

We simulated heavy-particle decays using $Z'$ bosons as they are included in the
PYTHIA model, forcing these states to decay to $t\bar{t}$ pairs. 
Both top quarks were set to decay  hadronically.
The PYTHIA parameters were set to the default ATLAS parameters tuned
to describe multiple interactions \cite{Moraes:2009zz}.  
The jets were reconstructed using the anti-$k_T$ algorithm
using the final-state hadrons.  
The events were first generated and stored for easy processing.

Figure~\ref{jetkt4a} shows the jet-shape variables 
for: (1) inclusive $pp$ production and (2) 
for $Z'$, with a mass of 2 and 3 TeV (solid and dashed lines, respectively). 
Events were selected with at least one jet with transverse momentum $p_T(jet)>500$ GeV and 
within the pseudorapidity range $|\eta(jet)|<2.5$. The histograms show the jet shape variables
using fully inclusive QCD events with an identical event selection. 

It is evident that all jet shapes are rather different for inclusive $pp$ collisions 
and collisions involving the production of TeV-scale particles decaying into $t\bar{t}$.
The observed 
differences between the two processes can easily be understood. 
Jets originating from top decays have masses 
close to the top-quark mass (for a complete overlap of decay products) 
or $W$ mass. Such jets are wider 
and are more elongated than standard QCD jets due to 
the decay plane formed by the channel  $t \to W b$ (eccentricity values are shifted to unity).

The jet mass for the second leading jet in $p_T$ is shown in Fig.~\ref{jetkt4b}.
The peak of this distribution is shifted towards the $W$ mass.
Other jet-shape variables are rather similar to those shown in Fig.~\ref{jetkt4a}.  

The use of the anti-$k_T$ algorithm with a large separation parameter $R$
is expected to be more efficient since
hadrons  from top decays  will have a higher chance to be  
contained inside single monojets. 
We have calculated the jet shapes 
for the anti-$k_T$ algorithm with $R=0.6$.
The largest difference with respect to the case $R=0.4$ 
is for the jet masses as shown in Fig.~\ref{jetkt6}.  
The $W$ mass peak 
shown in Fig.~\ref{jetkt4a}(a) is significantly reduced for jets with $R=0.6$.
At the same time, the peak attributed to the top quarks is more pronounced.
This is not surprising since a jet cone of 0.6 is more likely
to fully contain the decay products. 
The conclusion is that the anti-$k_T$ jets with $R=0.6$ are 
better suited for the searches using dijet invariant masses and the jet-shape variables,
since they allow a better separation of the signal events from the QCD-induced 
background. 
  
With the power of jet-shape variables in our hands, one can proceed with 
the rejection of QCD-induced jets in searches for  TeV-mass particles.
For simplicity,
we apply a simple cut method, i.e. 
we will place cuts on the  jet-shape variables to reject QCD background events. 
More elaborate studies will likely require the maximum-likelihood approach or a construction 
of a  neural network for event rejection. 
We also concentrate on the anti-$k_T$ algorithm with the size 0.4
as shown in Figs.~\ref{jetkt4a}-\ref{jetkt4b}.
The following selection cuts were used:

\begin{itemize}

\item
$M(jet) > 70\gev$, with $M(jet)$ being the jet mass; 

\item
$W(jet) > 0.07$, with $W(jet)$ being the jet width; 

\item
EEC$>0.75$.

\end{itemize}  
The above cuts are indicated on Fig.~\ref{jetkt4a}.
We applied identical cuts on the first and the second leading in $p_T$ jet,
although the cuts should be  optimized depending on the
average transverse momentum.
We also did not tune our cuts for the $3$ TeV mass range; these should
be somewhat looser in order to achieve  the best possible efficiency
for the reconstruction of $X$ particles.

It has already been discussed that $b$-tagging information can help in 
background reduction \cite{Lillie:2007yh}.
We did not use $b$-tagging information for our calculation, since its 
inclusion requires a detailed knowledge
of experimental efficiencies and purities. These will 
clearly be affected by the potentially large boost
of the $b$-quarks and thus a very detailed experimental study of this problem is necessary.

\begin{figure}[htp]
\begin{center}
\mbox{\epsfig{file=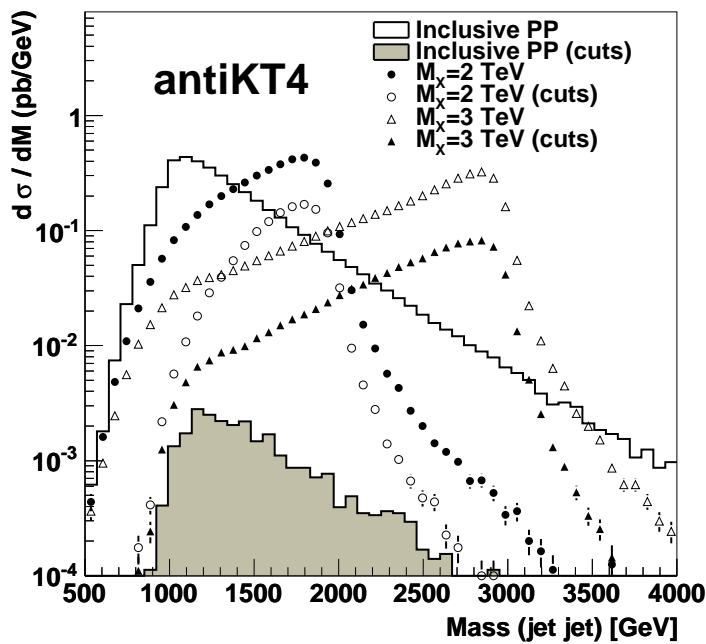,width=10cm}}
\caption
{
The jet-jet invariant mass before and after the applied event-selection cuts 
described in the text.
The signal events (before the selection cuts)
for particles with the masses 2 and 3 TeV are normalized to the background events.
}
\label{jetjet}
\end{center}
\end{figure}

Figure~\ref{jetjet} 
shows the jet-jet invariant mass before and after the applied cuts.
For illustration, the distribution with the signal events before the selection cuts
was normalized to the total number of background events.
The rejection factor for QCD events 
in the mass range $1.5-2$ TeV
is roughly 105, while it is only a factor of $2.7$ for the signal events.
Therefore, the ratio of the rejection factors for inclusive QCD and events with heavy states
is about 38.
A similar conclusion was obtained for 3 TeV particles,
although the rejection of the signal events was somewhat larger (3.7) since no attempt was made to fine
tune our cuts for this mass range.

\begin{figure}[htp]
\begin{center}
\mbox{\epsfig{file=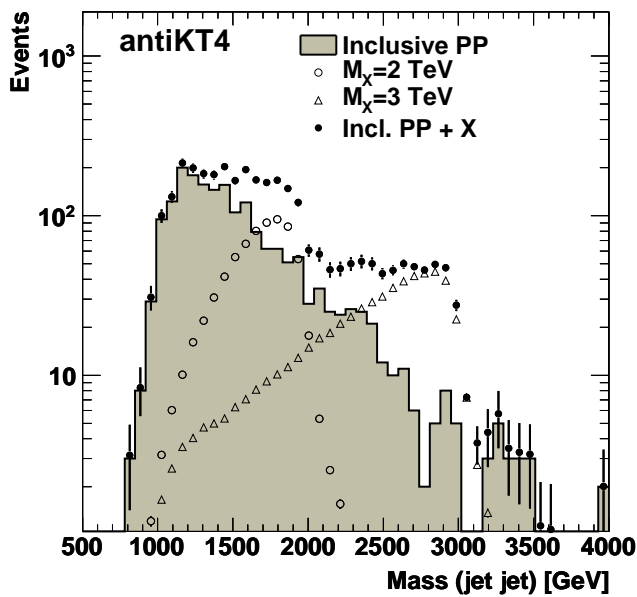,width=10cm}}
\caption
{
Jet-jet mass as in Fig.~\protect{\ref{jetjet}},
but the distribution is presented in terms of the 
number of events for 200 pb$^{-1}$ of integrated
luminosity.
Also shown is the fraction of signal events necessary for a $6~\sigma$
observation of a broad peak in the jet-jet invariant mass.
}
\label{jetjetsigma}
\end{center}
\end{figure}

Figure~\ref{jetjetsigma} shows the jet-jet invariant mass
for an integrated luminosity of 200 pb$^{-1}$  at 10 TeV center-of-mass
energy. The distribution is shown after the event-selection cuts 
described above.    
The shaded histogram shows the expected QCD 
background alone, while the solid symbols show QCD inclusive events plus
contributions of 2 TeV (open symbols) and  3 TeV (open triangles)
particles. The number of the signal events was determined   
in each case such that the overall signal-plus-background distribution features
two bumps with $6-7\sigma$ significance level.
The statistical significance of such enhancements was estimated using a $\chi^2$ fit 
with a Gaussian plus second-order
polynomial function. 

From the  
number of the signal events shown in Figure~\ref{jetjetsigma}, one can calculate  
the cross section $\sigma(X)  Br(X\to t\bar{t})$ 
for a discovery of a new TeV-mass particle at the $6\sigma$ confidence level.
For a 2 TeV mass, this cross section is  about $2-3$ pb.
This cross section can be as low as $1$ pb for 3 TeV particles.

In comparison with other discovery channels,
the obtained cross section for a $6\sigma$ observation 
is an order of magnitude smaller than that
expected for the SM
Higgs boson with the mass $150$ GeV, and roughly equals to the production cross
section for Higgs bosons with a mass 700 GeV \cite{Buttar:2006zd}.

\section{Discussion}

Although the goal of this paper is to extract the discovery-level cross section 
for $t\bar{t}$ resonances using jet-shape techniques,
it is natural to ask how the approach discussed in this paper is different from
other publications based on jet substructure.  

A rejection factor 100
for QCD inclusive events was achieved in Ref.~\cite{Kaplan:2008ie} using subjets 
in fully-hadronic $t\bar{t}$ decays and then imposing
kinematic constraints. The Cambridge-Aachen jet clustering 
algorithm \cite{Dokshitzer:1997in} was used in this approach which uses features 
of this algorithm to identify jets with substructure by decomposing jets into
subjets. 
A similar rejection factor was achieved  by the CMS 
Collaboration \cite{CMS-PAS-JME-09-001}
also using the Cambridge-Aachen jet clustering algorithm after decomposing jets
into subjets and examining kinematics of these subjets. Another closely related 
study has recently been performed by the ATLAS 
collaboration \cite{ATL-PHYS-PUB-2009-081} 
with the goal to improve the rejection efficiency discussed in their previous 
publication \cite{Brooijmans:2008,Brooijmans:2008se}.
While the jet substructure was used for the monojet, the main  improvement in
rejection of QCD background rejection comes from requiring semi-leptonic decays.   
   
The approach used in this paper does not include kinematic 
constraints on subjet kinematics, since we do not attempt to resolve kinematic characteristics 
of separate 
subjets inside jets. Instead, this paper relies primarily on the 
jet shapes (jet width and eccentricity) and 
jet masses. Thus the details of jet fine-structure are less important and, as a 
result, this method is expected to have smaller sensitivity to the fragmentation 
mechanism and to detector effects (such as calorimeter segmentation). Moreover, 
this approach is based on the anti-$k_{T}$ algorithm which is expected to have an 
advantage in events with pile-up.

\section{Summary}

We have shown that the jet-shape approach is extremely powerful 
for reduction of the contribution from QCD jets to dijet invariant masses
which can be used for searches of
TeV-scale particles. 
A rejection factor
above 100 can easily be achieved for inclusive $pp$ collisions,
with only a factor three reduction for the signal events for a 2 TeV particle 
decaying to $t\bar{t}$
using  jets reconstructed with the 
anti-$k_T$ algorithm with the size $R=0.4$.
This was obtained without sophisticated reconstruction methods, 
without careful tuning of cuts for both leading in $p_T$ jets
and without using $b$-tagging information.
This rejection factor is a factor of four larger compared to the case without using 
jet-shape information \cite{Lillie:2007yh} and similar to that obtained in \cite{CMS-PAS-JME-09-001}.

It should be noted that a full detector simulation may change the values 
of the rejection factors. 
However, no large change in our conclusions is expected 
since jets at the transverse momenta considered in this paper are typically  well reconstructed. 

The approach discussed in this paper is illustrated using  the
$X\to t\bar{t} \to W^+b_1 W^-b_2$
decay channel,   
with subsequent hadronic decays of the $W$ mesons.
Decay channels involving $t\bar{t}$ are the most promising since they involve top quarks
that are known couple most strongly to the electroweak symmetry breaking sector.
However, we expect a similar conclusion  for other cascade-type decays. 

The jet-shape approach allows to obtain a very competitive 
$\sigma(X) Br(X\to t\bar{t})\simeq 2-3$~pb cross section for a $6\sigma$ observation
of an enhancement in jet-jet invariant-mass spectra in the mass range 2-3 TeV.  
As example,
this cross section is one order of magnitude smaller than that 
expected for the SM Higgs production with the
mass 150 GeV at a similar center-of-mass energies.

\section*{Acknowledgements}
We would like to thank Brian Martin (MSU) and Gavin Salam (LPTHE) for useful discussions of the anti-$k_T$ algorithm 
and the FastJet package.

\bibliographystyle{./Macros/l4z_pl}
\def\bibname{\Large\bf References}
\def\refname{\Large\bf References}
\pagestyle{plain}
\bibliography{biblio}

\end{document}